\input{style/arxiv-general.cfg}
\documentclass[aoas,MSNbibl,nameyear,rotating,dvips]{arximspdf}
\makeatletter
   \@ifpackageloaded{graphicx}{}{\usepackage{graphicx}}
\makeatother

\doi{10.1214/14-AOAS796}
\volume{9}
\issue{1}
\pubyear{2015}
\firstpage{353}
\lastpage{382}
\docsubty{FLA}

\makeatletter
\def\mid{|}
\newcommand{\eqref}[1]{(\ref{#1})}
\makeatother

\begin{document}
\begin{frontmatter}

\title{Modeling for seasonal marked point processes:
An~analysis of evolving hurricane occurrences\thanksref{T1}}
\runtitle{Modeling for seasonal marked point processes}
\thankstext{T1}{Supported in part by the NSF under award SES 1024484.}

\begin{aug}
\author[A]{\fnms{Sai}~\snm{Xiao}\ead[label=e1]{sxiao@soe.ucsc.edu}},
\author[A]{\fnms{Athanasios}~\snm{Kottas}\corref{}\ead[label=e2]{thanos@soe.ucsc.edu}}
\and
\author[A]{\fnms{Bruno}~\snm{Sans\'o}\ead[label=e3]{bruno@soe.ucsc.edu}} %
\runauthor{S. Xiao, A. Kottas and B. Sans\'o}
\affiliation{University of California, Santa Cruz}
\address[A]{Department of Applied Mathematics and Statistics\\
University of California, Santa Cruz\\
Santa Cruz, California 95064\\
USA\\
\printead{e1}\\
\phantom{E-mail:\ }\printead*{e2}\\
\phantom{E-mail:\ }\printead*{e3}}
\end{aug}

%
\received{\smonth{3} \syear{2014}}
%
\revised{\smonth{10} \syear{2014}}

%
\begin{abstract}
Seasonal point processes refer to stochastic models for random
events which are only observed in a given season.
We develop nonparametric Bayesian methodology to study the dynamic
evolution of a seasonal marked point process intensity.
We assume the point process is a nonhomogeneous Poisson process
and propose a nonparametric mixture of beta densities to model
dynamically evolving temporal Poisson process intensities.
Dependence structure is built through a dependent
Dirichlet process prior for the seasonally-varying mixing distributions.
We extend the nonparametric model to incorporate time-varying marks,
resulting in flexible inference for both the seasonal point process
intensity and for the conditional mark distribution.
The motivating application involves the analysis of hurricane landfalls
with reported damages along the U.S. Gulf and Atlantic coasts from
1900 to 2010. We focus on studying the evolution of the intensity of
the process
of hurricane landfall occurrences,
and the respective maximum wind speed and associated damages.
Our results indicate an increase in the number of hurricane landfall
occurrences and a decrease in the median maximum wind speed
at the peak of the season. Introducing standardized damage as a mark,
such that reported damages are comparable both in time and space, we
find that there is no significant rising trend in hurricane damages
over time.
\end{abstract}

%
\end{frontmatter}


\section{Introduction}\label{sec1}

There are many examples of phenomena that occur every year at
random times but are limited to a specific season. Two examples of
natural events with strong scientific and economic relevance are the
following: the
Atlantic hurricanes and the Pacific typhoons formed by tropical
cyclones that occur between May and November; and the spawning of
coho salmon that takes place from November to January. There are some
situations where the observational window is limited to a given season,
such as wildlife abundance in regions that are not accessible in the
winter. In addition, there exist applications where interest lies in
studying a physical process during a particular season. One example
is the study of extreme precipitation during the dry season in tropical
environments. This can be important to guarantee water supplies and
also to prevent unexpected disasters. On a different note, studying
incidence of online purchase of products during the Christmas season is
indispensable for retailers in order to optimize stocking, advertising,
logistics, staffing, and website maintenance and support.
In all these examples it is
important to understand the underlying mechanism of the seasonal
point process. To this end, we need a flexible statistical model that
can describe
the changes of the process intensity during the season. The model also
has to capture the evolution of the intensities from one year to the
next, borrowing strength from the whole data set to improve the
estimation in a given season. Moreover, the model should be extensible to
allow for inference on possible marks associated with the occurrence of
the events.

In this paper, we focus on the study of landfalling hurricanes recorded
along the U.S. Gulf and Atlantic coasts between 1900 and 2010,
and their associated maximum wind speed and damages. Hurricanes are typical
seasonal extreme climate events. In light of potential societal and
economic impacts of climate change, the obvious
question regarding hurricanes is whether there is an intensification of
hurricane
frequency and an increasing trend of hurricane wind speed and
associated damage.
A substantial part of the literature on the variability of hurricane
occurrences is based on annual counts of events. For example,
\citet{Elsner2004} and \citet{Robbins2012} use change point
detection methods to find significant increases in storm
frequencies around 1960 and 1995. Limiting the analysis to the number
of hurricanes per year precludes the description of occurrence
variability within each year. Thus, it is not possible to estimate
trends in hurricane occurrence during a particular period within
the hurricane season, say, a given month. An alternative approach
is considered in \citet{ANZS:ANZS127} where the process of hurricane
occurrences is modeled with a continuous time-varying intensity
function within one year. However, in this case, the inter-annual
variability is not accounted for. An approach that models
intra-annual as well as inter-annual variability is presented
in \citet{solow1989}. The model is applied to a U.S. hurricane data set
(different from the one considered here) that consists of
monthly counts along the mid-Atlantic coast of the U.S. in 1942--1983.
The basic assumption is that the data correspond to a Poisson
process with a nonstationary intensity function. This is decomposed into
a secular and a seasonal component, estimated from annual and monthly
counts, respectively.
The analysis indicates no trend during the 1950s and a decreasing trend
in the 1970s for the secular component, and a stationary
seasonal cycle over time.

The focus on hurricane occurrence is of great importance in a
climatological context.
However, the frequency of hurricanes provides only a partial measure of
the threat
that these phenomena represent. When exploring the association of hurricane
strength with global warming, \citet{Emanuel2005} calls for research on
hurricane potential destructiveness. The disastrous impact to coastal
areas draws
the attention of the public, and government officials and policy makers need
reliable inferences on hurricanes' potential damage for
long-term action on economic development and population growth [\citet
{Pielke1997}].
For instance, in about ten years from Hurricane Fay in 2002 to
Hurricane Irene in 2011, hurricane landfalls have caused
around \$235 billion damages in 2013 values, and in 2005 Hurricane
Katrina alone caused more than \$80 billion in damage.
The devastation raises public concern about societal vulnerability to
extreme climate
[\citet{Katz2010}].

The statistical literature includes some work on exploring possible trends
in landfalling hurricanes' total damages. \citet{Katz2002} uses a
compound Poisson
process as a stochastic model for total damage. The model
consists of two separate components: one for annual hurricane
frequency, and a
second one for individual hurricane damage. The resulting analysis
suggests no upward trend for hurricane damages recorded between
1925--1995, after normalization due to societal changes. Damages are
modeled using a log-normal distribution and occurrences are assumed
to follow a homogeneous Poisson process, without any time-varying dynamics.
Moreover, the literature includes approaches that study the effect of
climate and physical factors on hurricane activity [\citet{ElsnerJagger2013}].
\citet{Katz2002} describes the association between hurricane damages
and El Ni\~{n}o. \citet{jagger2006climatology} apply extreme value
theory to hurricanes with extreme wind speeds. They assume a
homogeneous Poisson process for the occurrences of hurricanes
with wind speeds above a threshold, and a generalized Pareto distribution
for maximum wind speeds. They find that the quantiles of the
distribution of extreme wind speeds vary according to climate factors
that affect specific regions differently. Yet another association of
hurricane activity with climatic indexes is found in
\citet{JaggerElsnerBurch2011}, where hurricane damages are related
to the number of sunspots, as well as to the North Atlantic
Oscillation and the Southern Oscillation indexes.
\citet{chavas2012} model the damage index exceedance over a certain
threshold using the generalized Pareto distribution with several physical
covariates, such as maximum wind speed and continental slope.
\citet{MurnaneElsner2012} use quantile regression to study the
relationship between maximum wind speed and normalized economic losses.
Essentially, all the papers discussed above focus on estimating trends
in hurricane damage and/or its relationship with climate factors.
When the point process of hurricane occurrences is modeled, this is
done under the simplistic setting of a homogeneous Poisson process.

A fundamental question that remains unanswered by the previously
described work is whether the trend of hurricane damage over time is
due to the increasing/decreasing frequency or to more/less destructive
power of individual hurricanes. These are challenging questions, as
natural variability is large and we observe only a handful of
hurricanes per
season. These issues motivate the presentation of a new statistical
method for the analysis of the hurricane data.

In this paper, we propose a flexible joint model for inference on hurricane
frequency, maximum wind speed and hurricane damage. Our initial
assumption is that
the point process of hurricane landfalls follows a nonhomogeneous
Poisson process. As such,
the process is characterized by nonconstant intensity functions indexed
by the hurricane season. Notice that we refer to ``intensity'' using
the point
process terminology, and not the climate terminology, where it refers
to maximum
wind speed. We decompose the intensity functions into normalizing
constants, which
model annual hurricane frequencies, and density functions, which model
normalized
intensities within a season. We use a time series model for the normalizing
constants. We then take advantage of the flexibility of Bayesian nonparametric
methods to model the sequence of nonhomogeneous density functions. The proposed
approach allows for detailed inferences on both the intra-seasonal
variations of
hurricane occurrences, and the inter-seasonal changes of hurricane
frequencies. The
latter can be considered on time frames shorter than the whole season,
for example,
monthly. To our knowledge, this is the first statistical analysis of
hurricane behavior that takes such a comprehensive approach.
Moreover, to study hurricane damage, we treat maximum wind speed and
hurricane damage as marks associated with each hurricane occurrence. We
extend the method described above to make inference about marks
associated with
the time of occurrence of the point process events. As a result, we
obtain a full
probabilistic description of the dynamics of the process intensities and
the distribution of the marks. The application is focused on the
hurricane data, but the methodology is suitable in general for
time-varying seasonal marked Poisson processes.

The article is organized as follows. Section~\ref{section2} describes the hurricane
data and previous work relevant to this application. We perform an
initial analysis of the data, ignoring the year of hurricane occurrence
and using a mixture of Beta densities to model the hurricane intensity.
This analysis serves to motivate the methodological development, as it
clearly suggests that a simple parametric model would not capture
the complex shape of the intensity function of occurrences during the
hurricane season. Section~\ref{sec:intensity} develops the
methodology to incorporate dynamic evolution in the analysis, using
dependent Dirichlet process mixture models. We explore the problem
of data aggregation and study different aggregation strategies.
In Section~\ref{sec:mark} we present the extension
of the model to time-varying marks and apply it to maximum wind speed
and hurricane damage. Our results indicate that at the peak of the season,
there is an increase in the number of hurricane occurrences, a decrease
in the median maximum wind speed and a slight decreasing trend in
standardized damage associated with a particular hurricane.
Section~\ref{sec5} concludes with a general discussion.

%
%

\section{Hurricane data}
\label{section2}

We consider data for 239 hurricane landfalls with reported damages along
the U.S. Gulf and Atlantic coasts from 1900 to 2010. The data are
available from the ICAT Damage Estimator website
(\url{http://www.icatdamageestimator.com}).
ICAT provides property insurance to businesses and home owners
for hurricane and earthquake damage in the United States.
The ICAT data are consistent with the landfall summary data of the
National Hurricane
Center's North Atlantic hurricane database (HURDAT).
The scope of the data is restricted to landfalling hurricanes, as we
emphasize the analysis of a marked point process where damage is a mark
of key interest. Hurricanes are usually defined as tropical cyclones with
maximum wind speed of at least 74 miles per hour (mph).
With some abuse of terminology, we
use ``hurricanes'' throughout the paper to refer to all the storms in the
ICAT data set. This includes 4 tropical depressions, 63~tropical
storms, 54 hurricanes of category 1, 42 hurricanes of category 2, 59
hurricanes of category 3, 14 hurricanes of category 4, and 3 hurricanes of
category~5. The classification follows the Saffir--Simpson hurricane scale
in Table~\ref{tab:category}.
The data set includes information on the landing date, base damage,
normalized damage to current value, category, maximum wind speed and
affected states. A~detailed description of the data can be found in
\citet{Pielke2008} and the ICAT website. In particular, as discussed in
\citet{Pielke2008}, there is an undercount of damaging storms prior to 1940.
This is an important issue that needs to be considered when
quantifying possible trends in the number of hurricane occurrences.

\begin{table}
\tabcolsep=0pt
\caption{Saffir--Simpson hurricane scale. TD: tropical depression; TS:
tropical storm; HC 1 to HC 5: hurricane of category 1 to 5}\label{tab:category}
\begin{tabular*}{\textwidth}{@{\extracolsep{\fill}}lccccccc@{}}
\hline
\textbf{Category} & \textbf{TD} & \textbf{TS} & \textbf{HC 1} & \textbf{HC 2} & \textbf{HC 3} & \textbf{HC 4} & \multicolumn{1}{c@{}}{\textbf{HC 5}} \\
\hline
Maximum wind speed (mph) & $<$39 & 39--73 & 74--95 & 96--110 & 111--130 &
131--155 & $>$155 \\
\hline
\end{tabular*}
\end{table}

In this application, we consider maximum wind speed and economic
damage as marks. Maximum wind speed is defined as the
maximum sustained (over one minute) surface wind speed to occur along
the U.S. coast.
Economic damage is reported as base damage, which is the
direct total loss associated with the hurricane's impact in the year
when the hurricane occurred. In order to make all storm damages comparable,
a standardization method is used to estimate the damages to a baseline
year by extending the normalization method from \citet{Pielke2008};
see Section~\ref{sec:std_damage} for details.

\begin{figure}

\includegraphics{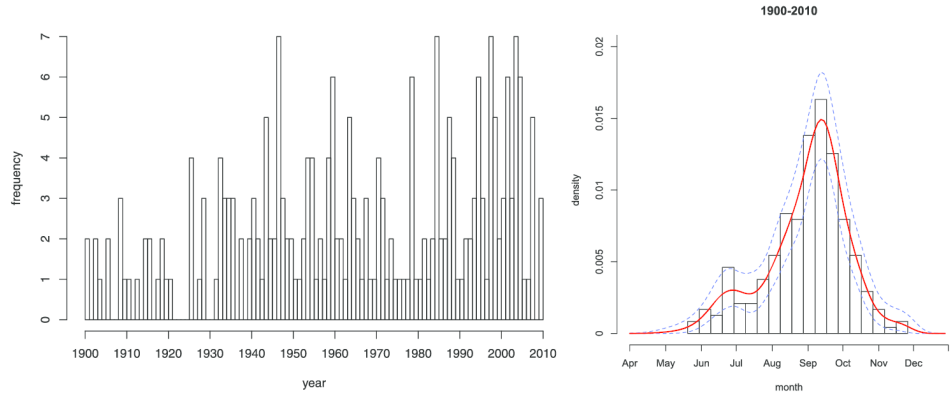}

\caption{Left panel: The time series of annual hurricane occurrences.
Right panel: Histogram (with bin width of 10 days) of hurricane occurrences
over months after aggregating all hurricanes into one year.
The solid and dashed lines denote the point and 95\% interval estimates
of the corresponding NHPP density, using the Dirichlet process mixture
model discussed in Section \protect\ref{section2}.}
\label{fig:hist_days}
\end{figure}

\begin{figure}

\includegraphics{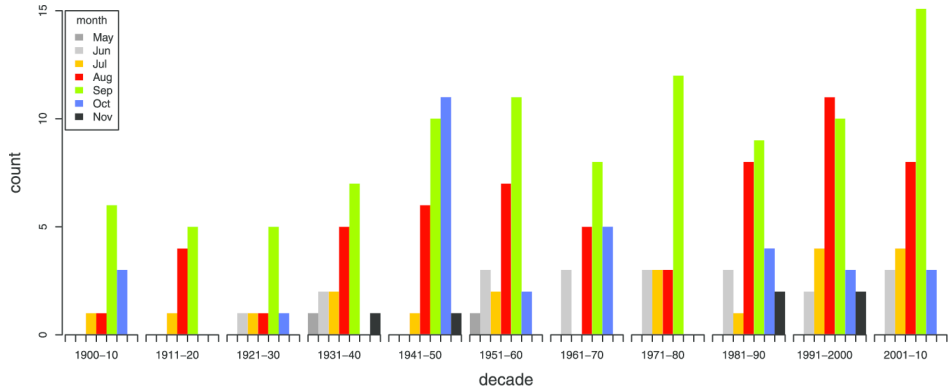}

\caption{The number of hurricanes within one season aggregated
by decades. In each decade, the number of hurricanes is grouped by months.}
\label{fig:hist_decade_month}
\end{figure}

The time series of annual hurricane counts is shown in
Figure~\ref{fig:hist_days}. Evidently, hurricane
occurrence depicts strong inter-annual variability. Moreover, there are
indications of discontinuities, which have been thoroughly considered in
the literature. In fact, significant shifts during the middle of the 1940s,
1960s and in 1995 have been identified in \citet{Elsner2004} and
\citet{Robbins2012}. The changes in the underlying data collection
methods, leading to change points in 1935 and 1960, have been
explained in \citet{Landsea1999} and \citet{Robbins2012}. To explore the
variability within the hurricane season, Figure~\ref{fig:hist_days}
also plots a histogram of hurricane occurrences ignoring the years of
the events.
The histogram reveals strong intra-seasonal variability, with the peak
of the
season around September and a concentration of hurricanes around June
during the early part of the season. Figure~\ref{fig:hist_decade_month}
provides further insight on the variability
of hurricane occurrence within the season, where we have now applied
aggregation by decades. The distribution of hurricane occurrences
within one season varies from decade to decade, and the
inter-decadal change of hurricane occurrences
varies from month to month. This indicates that the hurricane point
process intensity during a given season varies over the decades.
Here, we assume that such a process corresponds to a nonhomogeneous
Poisson process (NHPP).

There is a large body of literature on nonparametric methods to model
temporal (or spatial) NHPP intensities and to tackle the analytically
intractable NHPP likelihood.
Some are based on the log-Gaussian Cox process model [\citet{Moller1998},
\citet{Brix2001},
\citet{Liang2009}], while others use a Gaussian Cox process
model [\citet{adams-murray-mackay-2009c}].
An approach based on modeling the intensity function using kernel
mixtures of weighted gamma process priors is developed in
\citet{Wolpert1998} and \citet{Ishwaran2004}.
The method presented in this paper uses nonparametric mixtures to
model a density that, up to a scaling factor, defines the NHPP
intensity. The approach was originally developed in \citet{AK2006} and
\citet{Kottas2007}, with different applications considered by \citet
{ihler06b},
\citet{JiMerlKeplWest2009}, \citet{Taddy2010}, \citet{KBMPO2012} and
\citet{KWR2012}.

Let $\lambda(t)$ be the NHPP time-varying intensity, with $t$
in a bounded time window $(0, T)$. Inference proceeds by factoring the
intensity function as $\lambda(t) = \gamma f(t)$,
where $\gamma=\int_0^T\lambda(t)\,dt$ is the total
intensity over $(0,T)$; note that
$\gamma< \infty$ based on the local integrability of the NHPP
intensity function.
Hence, the likelihood function induced by the NHPP
assumption, using the observed point pattern $\{ t_1, \ldots, t_n\}$,
is given by
$p(\{ t_i \}_{i=1}^n | \gamma, f(\cdot)) \propto\exp(-\gamma) \gamma^{n} \prod_{i=1}^n f(t_i)$,
indicating that $f(t)$ and $\gamma$ can be modeled independently.
To develop inference for $\lambda(t)$, we start by rescaling all the
observations to the unit interval, thus setting $T=1$.
A convenient choice of distribution that will result in a conjugate
prior for $\gamma$ is the gamma distribution. Alternatively, we can
use the reference prior $p(\gamma) \propto\gamma^{-1} 1_{\{ \gamma>
0 \}}$
[\citet{AK2006}]. We model $f(t)$
using the density estimator given by the Dirichlet process (DP) mixture
model [\citet{Ferguson1973},
\citet{Antoniak1974}]. To complete the model we need
to specify a mixing kernel. The kernel of choice in this case is a Beta
density, which has the advantages of providing flexible shapes and,
being compatible with the compact support of the intensity, avoiding
edge effect problems. Using the DP stick-breaking representation
[\citet{Sethuraman1994}], the model can be formulated in the following
terms:
%
\begin{eqnarray}
\label{DP-mixture}  t_{i} \mid G, \tau&\sim& f(t_{i}\mid G,\tau)
=\int_{0}^{1} \operatorname{Beta}
\bigl(t_{i} \mid\mu\tau, (1-\mu)\tau\bigr) \,dG(\mu), \nonumber\\
 G (\mu) &= &\sum
_{j=1}^\infty w_j
\delta_{\mu
_{j}}(\mu),
\nonumber
\\[-8pt]
\\[-8pt]
\nonumber
 z_j &\stackrel{\mathrm{i.i.d.}} \sim&\operatorname{Beta}(1,
\alpha);\qquad w_1 = z_1, \\
 w_j &=& z_j
\prod_{r=1}^{j-1}( 1- z_r),\qquad j
\geq2; \qquad\mu_{j} \stackrel{\mathrm{i.i.d.}} \sim G_{0},\nonumber
\end{eqnarray}
where $G_{0}$ is the DP centering distribution and $\alpha$ is the DP
precision parameter. In our case, a convenient choice for $G_{0}$ is
given by
the uniform distribution noting that the Beta mixture kernel
is parameterized such that $\mu\in(0,1)$ is the mean and $\tau>0$
is a scale parameter.

We apply this model to the hurricane data ignoring the year
index. As shown in Figure~\ref{fig:hist_days}, the estimated density is
multi-modal, nonsymmetric and has a nonstandard right tail.
From this analysis it is clear that a proper description of the
hurricane data that assumes an underlying Poisson process requires a
nonhomogeneous intensity. Although the initial DP mixture model of
Beta densities is flexible enough to capture nonstandard shapes of
intensities within a season, it is not
capable of describing the evolution of intensities across seasons.
To address this problem, we propose in the next section a dynamic
extension of the Beta DP mixture model.

%
%

\section{Modeling time-varying intensities}
\label{sec:intensity}

We seek to model a collection of intensities evolving over years,
$\{ \lambda_{k}(t)\dvtx k\in\mathcal{K} \}$, where $\mathcal{K}=\{
1,2,\ldots \}$
denotes the discrete-time index set and $\lambda_{k}(t)$ is the
intensity for the season in year $k$. The model presented in the
previous section uses a DP prior to mix over the mean of
a Beta kernel. A temporal extension of such a model will have those priors
depend on $k$. To describe the correlation between successive years,
the model needs to impose dependence between the priors. As an
extension of the DP prior,
\citeauthor{Maceachern1999} (\citeyear{Maceachern1999,MacEachern2000})
proposed to model dependency across
several random probability measures. The extension is based on the
dependent Dirichlet process (DDP), which provides a natural way to model
data varying smoothly across temporal periods or spatial regions. The
construction of the DDP is based on the DP stick-breaking
definition, where the weights and/or atoms are replaced with appropriate
stochastic processes on $\mathcal{K}$. Here, we utilize the
``single-$p$'' DDP prior model, where the weights are constant over
$\mathcal{K}$, while the atoms are realizations of a stochastic process
on $\mathcal{K}$.

\subsection{Nonparametric dynamic model for Poisson process densities}
\label{sec:ddp}

Denote by $t_{i,k}$, for $i=1,\ldots,n_{k}$ and $k = 1,\ldots,K$,
the time of the $i$th event (hurricane landing date) in the $k$th
season, where
$K$ is the observed number of seasons and $n_k$ is the observation
count in the $k$th season. Recall that $ t_{i,k} $ has been
converted to the unit interval. Following the modeling approach
discussed in Section~\ref{section2}, the collection of NHPP intensities
can be represented by $\{ \lambda_k(t) = \gamma_k f_k(t)\dvtx k\in
\mathcal{K} \}$.
To introduce dependence on $\mathcal{K}$, we assume a parametric
time series model for $\{ \gamma_k\dvtx k\in\mathcal{K} \}$ and a DDP
mixture model for $\{ f_k(t)\dvtx k \in\mathcal{K} \}$. The former is
described in Section~\ref{sec:gamma}.
The latter is defined as follows:
\begin{eqnarray*}
f_{k}(t) &\equiv& f(t \mid G_{k},\tau) =\int
_0^1\operatorname{Beta}\bigl(t \mid\mu\tau, (1-
\mu)\tau\bigr)\,dG_k(\mu),\\
 G_k (\mu) &= &\sum
_{j=1}^\infty w_j \delta_{\mu_{j, k}}(
\mu),
\end{eqnarray*}
where the weights $\{ w_{j} \}$, defined as in (\ref{DP-mixture}), are
the same across seasons.
Thus, the model assumes that observations $t_{i,k}$
in the $k$th season arise from a mixture of Beta distributions with
component-specific means $\mu_{j,k}$ and variances
$\mu_{j,k} (1- \mu_{j,k}) / (\tau+ 1)$. The distribution for the mean
of the Beta mixture kernel is allowed to evolve over $\mathcal{K}$,
whereas $\tau$ is common to all $G_k$.

To impose dependence between the collection of random mixing
distributions~$G_k$, we replace $G_{0}$ in (\ref{DP-mixture}) with a
stochastic process for the atoms $\{ \mu_{j,k}\dvtx k \in\mathcal{K} \}$.
We thus need a discrete-time process with marginal distributions
supported on $(0,1)$, an appealing choice for which is the
positive correlated autoregressive process with Beta marginals
(PBAR) developed by \citet{Mckenzie1985}. For the atom $\mu_{j,k}$,
this is defined through latent random variables as follows:
%
\begin{equation}
\mu_{j,k} = v_{j,k} u_{j,k} \mu_{j, k-1} + (1-
v_{j,k}), \label{eq:one}
\end{equation}
where $\{v_{j,k}\dvtx k \in\mathcal{K} \}$ and $\{ u_{j,k}\dvtx k \in
\mathcal{K} \}$
are mutually independent sequences of i.i.d. Beta random variables,
specifically,
$v_{j,k} \sim\operatorname{Beta}(b, a-\rho)$ and $u_{j,k} \sim\operatorname
{Beta}(\rho, a-\rho)$,
with $a>0$, $b>0$ and $0 < \rho< a$. Using properties for products of
independent Beta random variables, it can be shown that (\ref{eq:one})
defines a stationary process $\{ \mu_{j,k}\dvtx k \in\mathcal{K} \}$ with
$\operatorname{Beta}(a,b)$ marginals. Moreover, the autocorrelation function
of the PBAR process is given by
$\{ \rho b a^{-1} (a + b - \rho)^{-1} \}^{m}$, $m=0,1,\ldots,$ and thus
$\rho$
controls the correlation structure of the process.

Although the DDP-PBAR prior for $G_{\mathcal{K}}=\{ G_k\dvtx k \in
\mathcal{K} \}$ is
centered around a stationary process, it generates nonstationary realizations.
In particular, if $\{ \theta_{k}\dvtx k \in\mathcal{K} \}$ given
$G_{\mathcal{K}}$ arises
from $G_{\mathcal{K}}$, then $\mathrm{E}(\theta_k \mid G_k) =\sum_{j=1}^\infty w_j \mu_{j,k}$ and
$\operatorname{Cov}(\theta_{k},\theta_{k+1} \mid G_{k},G_{k+1})=(\sum_{j=1}^\infty w_{j} \mu_{j,k} \mu_{j,k+1}) - (\sum_{j=1}^\infty w_j \mu_{j,k})
(\sum_{j=1}^\infty w_j\times\break  \mu_{j,k+1})$.

The Markov chain Monte Carlo (MCMC) method for inference, discussed in
Section~\ref{sec:inference-nomarks} and the \hyperref[app]{Appendix}, is based on a
truncation approximation to the DDP prior stick-breaking representation.
More specifically, $G_k \approx\sum_{j=1}^{N} w_j \delta_{\mu_{j, k}}$,
with $w_{1},\ldots,w_{N-1}$ defined as in (\ref{DP-mixture}), but $w_{N}=1-\sum_{j=1}^{N-1} w_{j}$. Because the weights are constant across seasons,
it is straightforward to choose the truncation level $N$ to any level
of accuracy using standard DP properties. For instance,
$\mathrm{E}(\sum_{j=1}^{N} w_{j} \mid\alpha)=1 - \{ \alpha/(\alpha+ 1) \}^{N}$, which can be averaged over the prior
for $\alpha$ to estimate $\mathrm{E}(\sum_{j=1}^{N} w_{j})$.
Given a tolerance level for the approximation, this expression can be
used to obtain the corresponding value $N$.
The truncated version of $G_{k}$ is used in all ensuing expressions
involving model properties and inference results.

\subsection{Time series model for the total intensities}
\label{sec:gamma}

The Poisson process integrated intensities $\{ \gamma_k \}$ can be
viewed as a realization from a time series in discrete index space, with
positive valued states. We adopt the state--space modeling method with
exact marginal likelihood proposed by \citet{Gamerman2013}. Unlike other
time series models that build from a log-Gaussian distributional
assumption, this approach provides a conjugate gamma prior,
resulting in an efficient MCMC algorithm for posterior simulation.
The model is defined by the following evolution equation for $\gamma_k$:
\[
\gamma_{k+1} = \frac{1}{\omega} \gamma_k
\xi_{k+1},\qquad \xi_{k+1} \mid\gamma_k, n_{1:k}
\sim\operatorname{Beta}\bigl(\omega a_k, (1-\omega) a_k
\bigr),
\]
where $\omega$ is a discount factor with $0 < \omega< 1$, $\xi
_{k+1}$ is a
random multiplicative shock,
and $n_{1:k}$ denotes the information available up to time $k$.

Denote $n_0$ as the information available initially. Take the initial
prior of $\gamma_0 \mid n_0$ as $\operatorname{Gamma}(a_0, b_0)$. Then, the prior
distribution at time $k$ is
$\gamma_k \mid n_{1:k} \sim\operatorname{Gamma}(a_{k|k-1}, b_{k|k-1})$, where
$a_{k|k-1} = \omega a_{k-1}$ and $b_{k|k-1} = \omega b_{k-1} $.
Based on the NHPP assumption,
$n_k \mid\gamma_{k} \sim\operatorname{Poisson}(\gamma_k)$, and thus the
updated distribution is $\gamma_k \mid n_{1:k} \sim\operatorname{Gamma}(a_k,b_k)$,
where $ a_k =\omega a_{k-1} + n_k$ and $b_k =\omega b_{k-1} + 1$.
The smoothing updated distribution is
%
\begin{equation}
\gamma_k - \omega\gamma_{k+1} \mid\gamma_{k+1},
n_{1:k} \sim\operatorname{Gamma}\bigl( (1 - \omega) a_k,
b_k\bigr). \label{smooth}
\end{equation}
For MCMC posterior inference, we can obtain samples from the full
conditionals of the
joint vector $\gamma_1, \ldots,\gamma_K$ by first filtering the
observations forward to obtain $a_k$ and $b_k$, $k=1, \ldots,K$, and
then sampling $\gamma_k$ backward, for $k=K, \ldots,1$, using the
distribution in \eqref{smooth}. The discount factor $\omega$ is estimated
by maximizing the joint log-likelihood function defined by the
observed predictive distribution $\log\prod_{k=1}^K p(n_k \mid
n_{1:k-1}, \omega) $.

\subsection{Implementation details and posterior inference}

\label{sec:inference-nomarks}

Inference for the scale parameter of the Beta mixture kernel using the fully
aggregated data (see Section~\ref{section2})
presented no problems and was quite robust to the choice of the gamma
prior assigned to $\tau$.
As discussed in more detail in Section~\ref{sec:results-nomarks}, to
estimate evolving hurricane intensities using the DDP mixture model,
it is necessary to apply some aggregation of the data into periods of
time that comprise more than one year. In this respect, aggregating
the data in decades emerges as an appropriate choice. However, the
estimation of $\tau$ becomes a challenging problem, since in
each decade there are still only a handful of hurricanes. In
fact, a simulation analysis indicates that reliable estimation of
$\tau$ requires between 50 to 100 observations per time period.
This problem can be explained by
the fact that $\tau$ partially controls the bandwidth of the Beta
kernels, with the width of the kernels in inverse relationship with the
size of $\tau$. Thus, when only a few data points are available, $\tau$
will tend to be small, allowing wide kernels to use the information from
most of the few available data. Such kernels cannot capture the
multi-modality of the seasonal hurricane intensity. We thus resort to fixing
the value of $\tau$ in our analysis of the data aggregated by decade. We
assume that the typical width of the Beta kernel corresponds to a month,
such that $(1/12)/4$ can be used as a proxy for the corresponding
standard deviation $\{ \hat{\mu}(1-\hat{\mu})/(\tau+ 1) \}^{1/2}$,
yielding $\tau=575$ when $\hat{\mu}= 0.5$. This is the value of
$\tau$
used in our analysis. We note that informative priors for $\tau$
centered around this value result in similar inferences.

For the centering PBAR process of the DDP prior, we set $a=b=1$,
leading to the default choice of uniform marginal distributions for
the $\mu_{j,k}$ covering the entire season between May and
November. The DDP prior specification is completed with a uniform
hyperprior for the PBAR correlation parameter $\rho$, and a
gamma$(2, 1)$ prior for $\alpha$.
Finally, we set $N=50$ for the truncation level in the DDP approximation;
note that under the gamma$(2, 1)$ prior for $\alpha$,
$\mathrm{E}(\sum_{j=1}^{50} w_{j}) \approx0.9999578$,
using the results discussed in Section~\ref{sec:ddp}.

We implement the DDP-PBAR model using the blocked Gibbs sampler
[\citet{Ishwaran2001}] with Metropolis--Hastings steps; see the \hyperref[app]{Appendix}
for details.
Combining the posterior samples for the parameters of the DDP-PBAR
model for $\{ f_k(t)\}$ and the posterior samples for the parameters of
the time series model for $\{ \gamma_k \} $, a variety
of inferences about hurricane intensity functionals can be obtained.

Of particular interest in our application is the average number of hurricanes
within a time interval $(t_1, t_2)$ in the $k$th season, which is
given by
$\Lambda_{k}(t_1, t_2) =\gamma_k \int_{t_1}^{t_2} f_k(t) \,dt$.
We can also obtain the probability of having a certain number $x$ of
hurricanes within time interval $(t_1, t_2)$ in the $k$th season as
$\{ (\Lambda_{k}(t_1, t_2))^{x}/x! \} \exp(-\Lambda_{k}(t_1, t_2))$.
As a consequence, the probability of having at least one hurricane
within time interval $(t_1, t_2)$ in the $k$th season is given by
$1 - \exp(-\Lambda_{k}(t_1, t_2))$. Under the DDP Beta mixture model,
$\int_{t_1}^{t_2} f_k(t) \,dt=\sum_{j=1}^N w_j \int_{t_1}^{t_2} \operatorname{Beta}(t \mid\mu_{j,k}
\tau, (1-\mu_{j,k}) \tau) \,dt$.

A further inferential objective is
the one-step ahead prediction of the intensity function for the next season,
$\gamma_{k+1} \sum_{j=1}^{N} w_{j} \operatorname{Beta}(t \mid\tilde{\mu
}_{j,k+1} \tau,
(1 - \tilde{\mu}_{j,k+1}) \tau)$. Based on the PBAR construction in
(\ref{eq:one}), the conditional distribution for $\tilde{\mu}_{j,k+1}$
given $\mu_{j,k}$ and $v_{j,k+1}$ is a rescaled version of the
$\operatorname{Beta}(\rho, 1-\rho)$ distribution for $u_{j,k+1}$. Hence, for each
$j=1,\ldots,N$, posterior predictive samples for the $\tilde{\mu}_{j,k+1}$
can be readily obtained given draws for the $\mu_{j,k}$ and $v_{j,k+1}$;
the former are imputed in the course of the MCMC, the latter can be
sampled from their $\operatorname{Beta}(1,1-\rho)$ distribution given the
MCMC draws for $\rho$. Therefore, combining with predictive draws for
$\gamma_{k+1}$, full inference is available for forecasting any
functional of the hurricane intensity.

\subsection{Analysis of dynamically evolving hurricane intensities}
\label{sec:results-nomarks}

\subsubsection{Data aggregation}

The number of landfalling hurricanes with reported damages during 1900--2010
in the U.S. is 239. On average, there are merely 2 or 3 hurricanes
every year, with no hurricane in some years, for example, 1922--1925 and 2009.
Thus, the first practical problem we face is that of data scarcity.
When modeling the data at the yearly level, the challenge is that it is
difficult to analyze a process with so few realizations per year.
Hence, we consider aggregating the data over periods of five and ten
years, and compare the results under the two different levels of aggregation.

\begin{figure}

\includegraphics{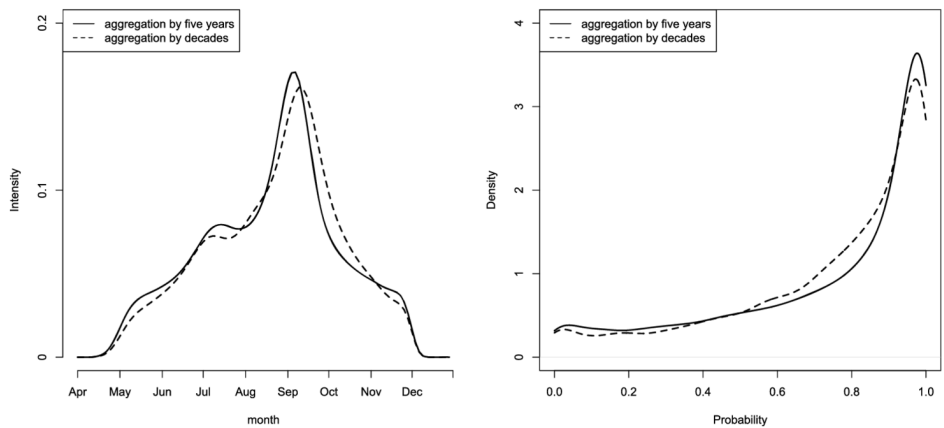}

\caption{Under the two distinct levels of data aggregation, posterior
mean estimates for the hurricane intensity in 2011--2015 (left panel)
and posterior densities for the probability of at least one hurricane
in May for 2011--2015 (right panel).}
\label{fig:compare1}
\end{figure}

Aggregation over a period of time is based on the assumption that the
NHPP densities for all the years corresponding to the aggregated period are
the same. For the five year aggregation we have 22 different intensities
and for the decadal aggregation we have 11. Data aggregation does not
effect the estimation of normalizing constants $\{ \gamma_k \}$. In
fact, we can apply the model for the $\{ \gamma_k \}$ proposed in Section~\ref{sec:gamma} to the yearly data, and then aggregate.
Figure~\ref{fig:compare1} provides results to compare the two aggregation
strategies in the context of forecasting the hurricane intensity and
one of its
functionals in the next five years 2011--2015. Encouragingly, the
results are very similar under the two levels of data aggregation.

\begin{figure}[b]

\includegraphics{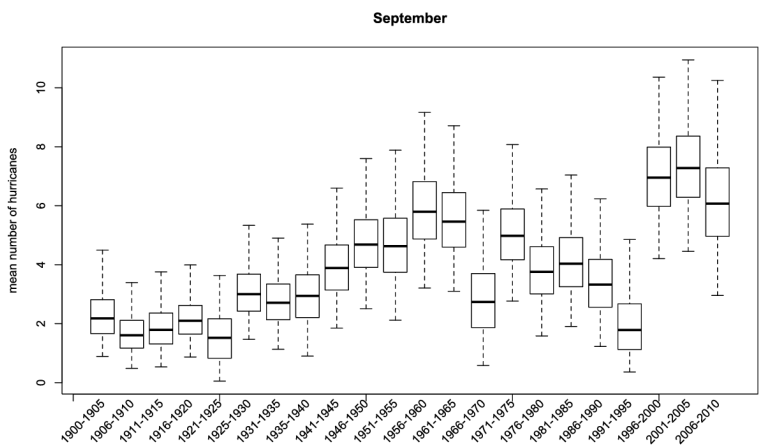}

\caption{Boxplots of posterior samples for the average number of
hurricanes in
the month of September across five-year periods from 1900 to 2010.}
\label{fig:compare2}
\end{figure}

\begin{figure}

\includegraphics{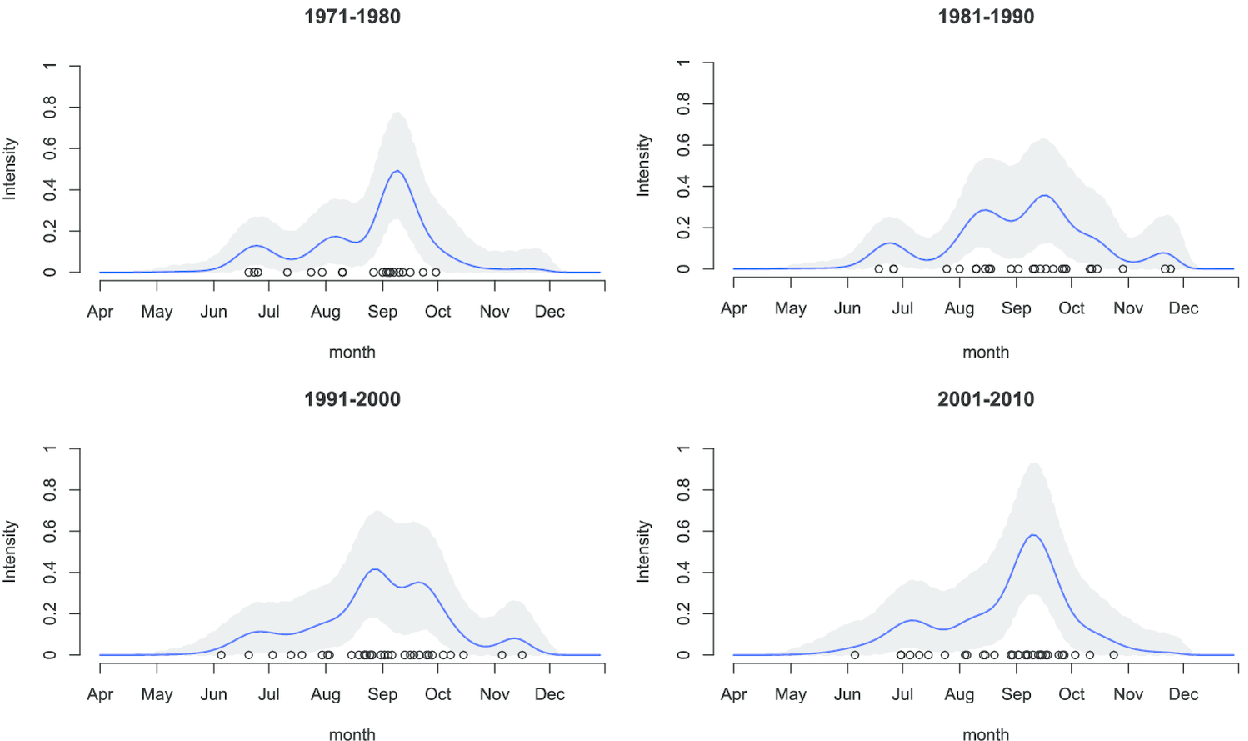}

\caption{Posterior mean estimates (solid line) and 95\% intervals
(gray bands) of the hurricane intensity during 1971--2010.
Points along the horizontal axis correspond to the observations.}
\label{fig:decade8-11}
\end{figure}

Regarding the analysis of historical data, we focus on the month of
September. In fact, for the Atlantic hurricane season, August, September
and October (ASO) are very important months, as 95\% of Saffir--Simpson
category 3, 4 and 5 hurricane activity occurs during August to October
[\citet{Landsea1993}]. In particular, September is the most
frequently occurring month. Figure~\ref{fig:compare2} shows the
estimated average number of hurricanes in September under the five year
data aggregation. We observe a strong variability, in particular, for the
periods 1921--1925, 1966--1970 and 1991--1995.
This can be attributed to the fact that during 1921--1925 there was no
hurricane in September. Moreover,
there was only one hurricane in September during 1966--1970, but
there were 7 hurricanes in September during both 1961--1965 and 1971--1975.
Finally, there was no hurricane in September during 1991--1995,
but 10 hurricanes occurred in September during 1996--2000.
Thus, even though the prior model is imposing some smoothness, posterior
inference results are still strongly affected by the scarcity of
observations, even at the level of a five year period.
Our resulting inference in the five-year aggregation level reflects the
strong variability of hurricane counts in September. More specifically,
the clear separation of the posterior distributions for the different
periods mentioned above gives a probabilistic assessment of significant
breakpoints. These are in agreement with the change points detected in
\citet{Elsner2004} and \citet{Robbins2012} for the counts over all months.
However, in this work we focus on revealing possible long-term trends
rather than on anomaly detection. Thus, on the basis of these analyses,
for the rest of the paper we focus on data aggregated over decades.

\begin{figure}

\includegraphics{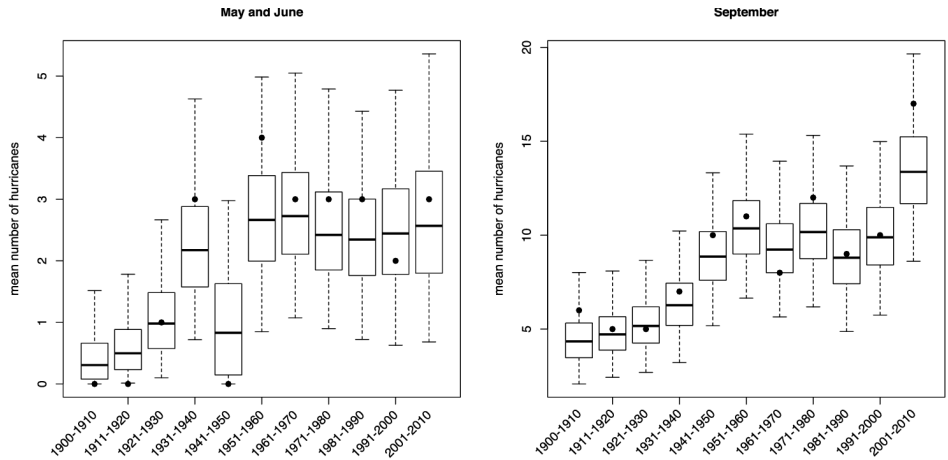}

\caption{Boxplots of posterior samples of the mean number of hurricanes
in early season (May and June) by decade (left panel) and in
September by decade (right panel). In both panels, the solid dots
indicate the corresponding observed numbers of hurricanes.}
\label{fig:boxplot}
\end{figure}

\subsubsection{Evolving hurricane intensities across decades}

Figure~\ref{fig:decade8-11} presents the estimated intensity
functions in the most recent four decades. The estimates fit the data
very well,
correctly capturing the peaks in ASO and tails in June and
November. They show some similarities between the decades, but they
adapt to the characteristic of the distribution of hurricane events in
each decade. An
important product of our probabilistic analysis is
the average number of hurricanes in a given
time period, which, as discussed in Section~\ref{sec:inference-nomarks},
requires the posterior distribution for both $\gamma_k$ and $G_k$.
In Figure~\ref{fig:boxplot} we present the distributions for the
mean number of hurricanes in the peak month of September and the
off-season months of May and June, along with the associated
observed number of hurricanes. Inference based on our model
smooths the data through the decades, especially when a
small number of observations are available. Overall, the
distribution of the mean number of hurricanes in each decade
matches the observations quite well.
Both panels depict an increasing trend in the first four decades as
well as during the most recent three decades. The former may be an
artifact of the under-reporting during the beginning of the 20th Century.
While the latter is very subtle for the off-season months, it is very strong
for the month of September. In fact, the last decade depicts an average
number of hurricanes in the peak of the season, which is substantially
higher than any other decade on record.

%
%

\section{DDP model for seasonal marked Poisson processes}
\label{sec:mark}
Here, we extend the DDP model, developed in the previous
section, to a seasonal marked Poisson process.
A marked Poisson process (MPP) refers to a Poisson process with an
associated random variable or vector for each event. In our
application, $\{ t_{i,k}\dvtx i=1,\ldots,n_k \}$ is a point pattern on $(0,
T)$ and the
marks can be denoted as $\{ y_{i,k}\dvtx i=1,\ldots,n_k \}$ on mark space $Y$.
Thus, the realization from the marked point process in the $k$th
decade is
$\{ (t_{i,k}, y_{i,k})\dvtx t _{i,k} \in(0, T), y_{i,k} \in Y\}$.
A MPP can be defined as a Poisson process on the joint marks-points
space with intensity function $\varphi$ on $(0,T) \times Y$.
In particular, the marking theorem [\citet{Moller2004}] states that
a MPP is a NHPP with intensity function given by $\varphi(t,y)=\lambda(t) f(y \mid t)$, where $\lambda(t)$ is the marginal temporal
intensity function, and the conditional mark density $f(y \mid t)$ depends
only on the current time point $t$.

\subsection{The DDP-AR model}
\label{sec:model_with_marks}

We extend the methodology from \citet{TaddyKottas2012} for MPPs based
on joint mixture modeling on the marks-points space. This modeling
approach yields flexible inference for both the marginal temporal
intensity and for the conditional mark distribution. Here, it is
utilized to develop a model for the collection of hurricane MPPs
evolving over decades. We will refer to the full model as the DDP-AR model,
since, in addition to the PBAR structure, it incorporates autoregressive
processes to model the conditional evolution of marks over time.

The marks are given by the maximum wind speed for each hurricane
and the associated economic damages.
Instead of using the total dollar amount of hurricane damage, we
define a standardized damage, which is calculated as a proportion of total
wealth with respect to a reference region and a baseline year (see
Section~\ref{sec:std_damage}). The resulting NHPP is defined in
a three-dimensional space comprising time, maximum wind speed and
standardized damage. Maximum wind speed and standardized damage are
transformed by taking logarithms and subtracting the global average
of the log-transformed values.
We denote $y_{i,k}$ and $z_{i,k}$ as, respectively, the
transformed maximum wind speed and the transformed standardized damage
of the $i$th hurricane in the $k$th decade. For the
three-dimensional intensity function, $\varphi_{k}(t, y, z)$, we use the
factorization $\gamma_k f_k(t, y, z)$, where $\{ \gamma_k \}$ follows the
time series model presented in Section~\ref{sec:gamma}. Regarding the
density function, we use a DDP mixture with a product of univariate
kernel densities for time and marks. Thus, the dependence among time
and marks is
introduced by the mixing distribution. We retain the Beta kernel
density for time and use Gaussian kernel densities on the log scale
for the two marks, mixing on the mean of each kernel component. Hence,
the DDP mixture model for $f_k(t, y, z)$ can be expressed as
%
\begin{equation}
\label{eq:joint-intensity} \int\operatorname{Beta}\bigl(t \mid\mu\tau,(1-\mu) \tau\bigr)
\mathrm{N}\bigl(y \mid\nu,\sigma^2\bigr) \mathrm{N}\bigl(z \mid\eta,
\zeta^2\bigr) \,dG_k(\mu, \nu, \eta),
\end{equation}
where $G_k (\mu, \nu, \eta)=\sum_{j=1}^{N} w_j \delta_{(\mu_{j,k},\nu_{j,k}, \eta_{j,k} )}
(\mu,\nu,\eta)$.
The locations $\nu$ and $\eta$ of the normal kernels are allowed to
change across decades. The scales $\sigma^2$ and $\zeta^2$ are
the same across decades, serving as adjusting parameters for the
bandwidth of the kernels. Conditionally conjugate inverse gamma priors
are assumed for $\sigma^2$ and $\zeta^2$.

Dependence across decades for maximum wind speeds and
standardized damages is obtained through AR(1) processes for
the respective kernel means $\{ \nu_{j,k}\dvtx k \in\mathcal{K} \}$ and
$\{ \eta_{j,k}\dvtx k \in\mathcal{K} \}$:
\[
\nu_{j,k} \mid\nu_{j, k-1} \sim\mathrm{N}\bigl(\beta
\nu_{j, k-1}, \sigma^2_1\bigr),\qquad \eta_{j,k}
\mid\eta_{j, k-1} \sim\mathrm{N}\bigl(\phi\eta_{j, k-1},
\sigma^2_2\bigr)
\]
with inverse gamma priors assigned to $\sigma_{1}^{2}$ and $\sigma_{2}^{2}$,
and uniform priors on $(-1,1)$ placed on $\beta$ and $\phi$. Since the
DDP prior structure for $G_{\mathcal{K}}=\{ G_k\dvtx k \in\mathcal{K}
\}$ in
(\ref{eq:joint-intensity}) extends the one for the DDP-PBAR model, we
retain the result about nonstationary realizations given $G_{\mathcal{K}}$,
extending the argument in Section~\ref{sec:ddp}.
When the random measures $G_k$ are integrated out, we obtain
$\mathrm{E}(y_k) = 0$, $\operatorname{Var}(y_k)=\mathrm{E}(\sigma^2) +
( 1- \beta^2)^{-1} \mathrm{E}(\sigma^2_1)$ and $\operatorname{Cov}(y_k,y_{k+1})=\beta(1- \beta^2)^{-1} \mathrm{E}( \sigma^2_1)$, with analogous
results for the $z_{k}$. These expressions can be of help for prior
specification.

The MCMC method for the DDP-AR model involves an extension of the
posterior simulation algorithm described in the \hyperref[app]{Appendix}.\setcounter{footnote}{1}\footnote{The code to implement the DDP-AR model (as well as the DDP-PBAR model)
is available from the first author's website at
\url{http://users.soe.ucsc.edu/\textasciitilde sxiao/research.html\#software}.}
As the marks are associated with normal AR(1)
processes and conditionally conjugate priors are used, all the
parameters associated
with marks have closed-form full conditionals. Finally, since the normalizing
factors (required for the standardization of damages) corresponding to
the period 2005--2010 are not available, the MCMC algorithm includes
steps to impute the missing standardized damages for those years.

\subsection{Standardization of hurricane damages}
\label{sec:std_damage}

The purpose of standardizing hurricane damages is to isolate
societal and spatial factors that affect the amount of damage and
are not considered in the model. There exist several methods to adjust
the economic damages of past hurricanes to today's value
[\citet{Pielke2008},
\citet{Schmidt2010},
\citet{Collins2001}]. Here, we define
standardized damage as an extension to the method in
\citet{Pielke2008}.

The hurricane data set includes base damage and normalized damage.
Base damage is calculated as the total landfall year dollar value
of the damage caused by a hurricane. Such amount is converted to
the dollar value corresponding to the latest year in the record by
normalizing for inflation, wealth and population over time. Denote
inflation, wealth per capita and affected county population in year $t$
as $I_{t}$, $W_{t}$ and $P_{t}$, respectively. Equation
\eqref{eqn:normalization} shows the normalization of the damage due to a
hurricane landing in year $t$ to values in year $s$:
%
\begin{equation}
\label{eqn:normalization} \mathrm{normalized.damage}_{s} = \mathrm{base.damage}_t
\times \frac{I_{s}}{I_{t}} \times\frac{W_{s}}{W_{t}} \times \frac{P_{s}}{P_{t}}.
\end{equation}
This normalization method yields the estimated damages of all hurricanes
in today's value but in the same region, for example, the damages
caused by
Katrina 2005 if it occurred under societal conditions in Louisiana
affected counties in 2013.

To make hurricane damages comparable, we have to
adjust for inflation and account for the fact that
much more damage will be caused if the hurricane lands in densely
populated and wealthier counties than in scarcely populated and poor
regions. Thus, we have to remove both a spatial and societal
factor from the damage, so that the model can explore
the pure association between damages and climate variability.
Hence, we define standardized damage as
\[
\mathrm{standardized.damage} = \frac{\mathrm{base.damage}_{t}}{I_{t}
\cdot
W_{t} \cdot P_{t}}.
\]
Such a quantity can be interpreted as a base damage normalized to a reference
year's value in a reference region; in the reference year and region, the
inflation factor, wealth per capita and population are all equal to 1.
This method removes the difference in hurricane damages due to the
landing years and locations. \citet{Eric2011}\vadjust{\goodbreak} and \citet{chavas2012}
developed similar ideas normalizing damages by using
$\mathrm{base.damage}_{t} / \mathrm{wealth}_{t}$, where
$\mathrm{wealth}_{t}$ is the total wealth of the affected regions. They
interpret the standardized damage as a relative damage, termed
actual-to-potential-loss ratio. Note that the denominator we use,
$I_{t} \cdot W_{t} \cdot P_{t}$, is an approximation of
$\mathrm{wealth}_{t}$. All inferences presented in Section~\ref{full-model-results}
that involve hurricane damage refer to standardized damage.
Note that, if the normalizing factors are provided, actual hurricane
damages for a given affected region and year can be obtained
from standardized damages.
It is important to notice that the normalizing factors prior to 1925
have larger
uncertainties compared to those for later periods
[\citet{Pielke2008}]. This problem is compounded with the already
mentioned issue of underreporting of hurricanes in the early part of the
20th Century. The reader should keep this in mind when interpreting
the results in the following sections.

\begin{figure}

\includegraphics{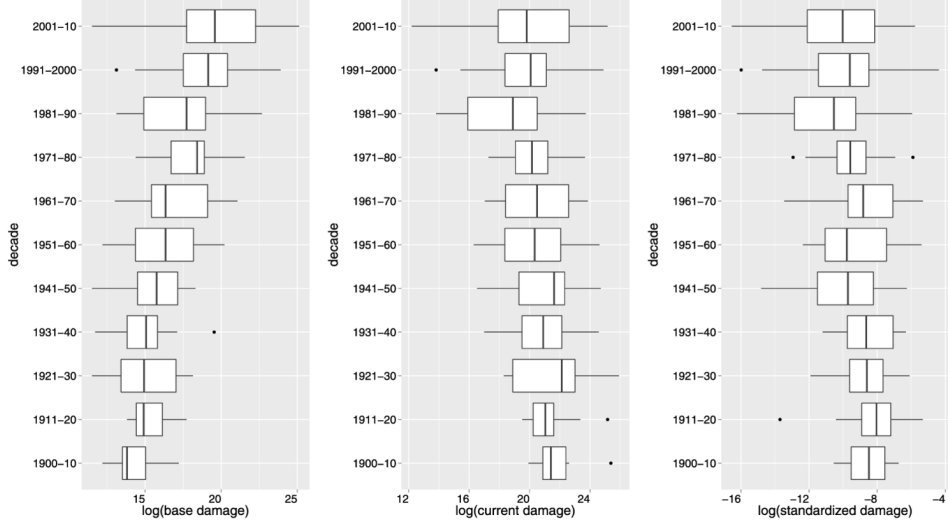}

\caption{Data box plots across decades for log-transformed base damages
(left panel), damages normalized to current values (middle panel) and
standardized damages (right panel).}
\label{fig:damage_conversion}
\end{figure}

To visualize the effect of the conversion on damage values, Figure~\ref{fig:damage_conversion} shows three different calculations for
hurricane damage and their change over decades.
The base damage depicts an increasing trend over decades, which
disappears after normalization and standardization.

\subsection{Inference}
\label{sec:inference}

For a marked point process the typical inference of interest is for the
distribution of the marks, conditional on time. To obtain inference about
different functionals of the conditional mark distribution, we use
the available posterior samples of the joint density $f_k(t, y, z)$.
Specifically,\ conditional inference for maximum wind speed is obtained from
%
\begin{eqnarray}
\label{eqn:density1} f_k( y \mid t, G_k) & =&
\frac{f_k (y, t \mid G_k)}{f_k (t \mid G_k)} \nonumber\\
&=& \frac{ \sum_{j=1}^{N} w_j \operatorname{Beta}(t \mid\mu_{j,k} \tau,
(1-\mu_{j,k}) \tau)
\mathrm{N} (y \mid\nu_{j,k}, \sigma^2) }{\sum_{j=1}^{N} w_{j}
\operatorname{Beta}(t \mid\mu_{j,k} \tau, (1-\mu_{j,k}) \tau) }\\
& = &\sum_{j=1}^{N}
w^*_{j,k}(t) \mathrm{N} \bigl(y \mid\nu_{j,k},
\sigma^2\bigr),\nonumber
\end{eqnarray}
where $w^*_{j,k}(t)=\frac{w_j \operatorname{Beta}(t \mid\mu_{j,k} \tau,
(1-\mu_{j,k}) \tau)
}{\sum_{j=1}^{N} w_j \operatorname{Beta}(t \mid\mu_{j,k} \tau, (1-\mu
_{j,k}) \tau)}$.
Of particular importance is the distribution of maximum wind speed
conditional on a specific time period, for example, the peak season ASO
or a
particular month. Suppose that the time period of interest corresponds to
the interval $(t_1, t_2)$. The density conditional on $(t_1, t_2)$
can be developed as
%
\begin{eqnarray}
\label{eqn:density2}&& f_k\bigl( y_{0} \mid t
\in(t_1, t_2), G_k\bigr) \nonumber\\
&& \qquad =
\frac{ \lim_{\Delta y_{0}
\to0}
({1}/{\Delta y_{0}})
\operatorname{Pr}(y \in(y_{0}, y_{0} +\Delta y_{0}], t \in(t_1, t_2) \mid
G_{k}) }{
 \operatorname{Pr}(t \in(t_1, t_2) \mid G_k) }\\
  &&\qquad=\sum_{j=1}^{N}
h^*_{j,k} \mathrm{N}\bigl(y_{0} \mid\nu_{j,k},
\sigma^2\bigr),\nonumber
\end{eqnarray}
where $h^*_{j,k} \equiv h^*_{j,k}(t_{1},t_{2}) =\frac{w_j \int_{t_1}^{t_2}
\operatorname{Beta}(t \mid\mu_{j,k} \tau, (1-\mu_{j,k}) \tau)
 \mathrm{d}t }{\sum_{j=1}^{N} w_j \int_{t_1}^{t_2}
\operatorname{Beta}(t \mid\mu_{j,k} \tau, (1-\mu_{j,k}) \tau)
 \mathrm{d}t}$.

In equations \eqref{eqn:density1} and \eqref{eqn:density2}, both the
weights, $w^*_{j,k}(t)$, $h^*_{j,k}$, and the mixing
components, $\nu_{j,k}$, change with the decade index $k$; importantly,
the former are time dependent, thus allowing local learning under the
implied location normal mixtures. Hence, the model has the flexibility
to capture general shapes for the conditional mark distribution which
are allowed to change across decades in a nonstandard fashion.
Analogous expressions hold for the conditional distribution of
standardized damage. Moreover, since equation
\eqref{eq:joint-intensity} provides the joint density of
time, maximum wind speed and standardized damage, we can obtain
inference for a mark conditional on an interval of the other mark and
an interval of time. For instance, we can explore the distribution of
damage conditional on the hurricane category as defined by different
intervals of maximum wind speed; see Table~\ref{tab:category}.

The time evolution of hurricane occurrences and the marks are
controlled by
autoregressive processes. One-step ahead prediction of joint time-mark
distributions can be obtained by extending the method described in
Section~\ref{sec:inference-nomarks} with additional sampling for the
$\{ \nu_{j,k+1} \}$ and $\{ \eta_{j,k+1} \}$ from the AR(1) processes
that form the building blocks of the DDP prior.


\subsection{Results}
\label{full-model-results}

We applied the DDP-AR model to the full
data set involving hurricane occurrences across decades and the associated
maximum wind speeds and standardized damages. The hyperpriors for the
time component of the DDP mixture model were similar to the ones
discussed in Section~\ref{sec:inference-nomarks} for the DDP-PBAR
model; $\tau$ was again fixed. For the variances of the Gaussian
mixture kernels and the variances of the corresponding AR(1) processes
for the DDP prior, we used $\sigma^2 \sim\operatorname{IG}(3, 2)$,
$\zeta^2 \sim\operatorname{IG}(3, 10)$ and $\sigma^2_1 \sim\operatorname{IG}(3, 2)$,
$\sigma_2^2 \sim\operatorname{IG}(3,10)$. Here, the shape parameter of each
inverse gamma prior is set to 3, which is the smallest (integer) value that
ensures finite prior variance. The prior means were specified using
the expressions for the marginal variances of maximum wind
speed and standardized damage (see Section~\ref{sec:model_with_marks})
with $\beta$ and $\phi$ replaced by their prior mean at 0. In
particular, we set $\mathrm{E}(\sigma^2)= \mathrm{E}(\sigma_1^2)= 0.5 (R_{y}/4)^{2}$
and $\mathrm{E}(\zeta^2)= \mathrm{E}(\sigma_2^2)= 0.5
(R_{z}/4)^{2}$, where $R_{y}$
and $R_{z}$ denotes the range of the $y_{i,k}$ and $z_{i,k}$, respectively.

\begin{figure}

\includegraphics{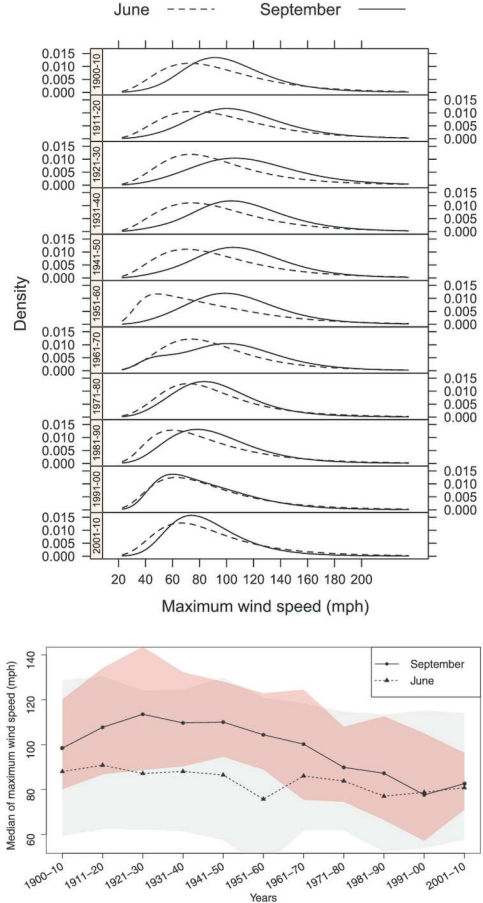}

\caption{Top panel: maximum wind speed densities conditional on June and
September for all decades. Bottom panel: posterior expectation and
95\% interval (dark gray band for September; light gray for June) for
the median
maximum wind speed in June and September versus decade.}
\label{fig:winds_month}
\end{figure}

The posterior distribution for the number of distinct mixing
components is supported by values that range from 10 to 16.
The 95\% posterior credible interval for $\rho$ is given by $(0.73,0.87)$,
resulting in a $(0.59,0.79)$ 95\% credible interval for the PBAR correlation.
On the other hand, the 95\% posterior credible intervals for $\beta$
and $\phi$ are, respectively, $(-0.14,0.79)$ and $(-0.24,0.81)$,
indicating more variability in the estimated correlation of the AR(1)
centering processes for the DDP prior.
Retaining the uniform priors for $\rho$, $\beta$ and $\phi$, we
performed a prior sensitivity analysis for the variance hyperparameters.
The parameters $\sigma^2$ and $\sigma_1^2$ associated with maximum
wind speed are relatively sensitive to the prior choice, while the parameters
$\zeta^2$ and $\sigma_2^2$ for standardized damage are quite stable.
Overall, posterior inference results are robust to moderate changes in the
prior hyperparameters.

For inference, we focus on the densities of maximum wind speed
and logarithmic standardized damage conditional on events occurring
in the early season and the peak season.
Figure~\ref{fig:winds_month} shows the comparison between
the maximum wind speed densities conditional on June and
September in each decade. We observe that maximum wind speeds in
September are higher than in June for all decades. In the 1960s the
density has a very long left-hand tail, even showing evidence of two
modes. Noteworthy in the last four decades is the increasing
accumulation of
density on lower values of maximum wind speed. The fact that maximum
wind speeds in September are decreasing is confirmed by the plot in the
lower panel of Figure~\ref{fig:winds_month}, where both point and interval
estimates support a decreasing trend for the median maximum wind speed
in September. In particular, after peaking at more than 110 mph in the
1920s, the posterior point estimate has settled at around 85 mph in
the last decade.

\begin{figure}

\includegraphics{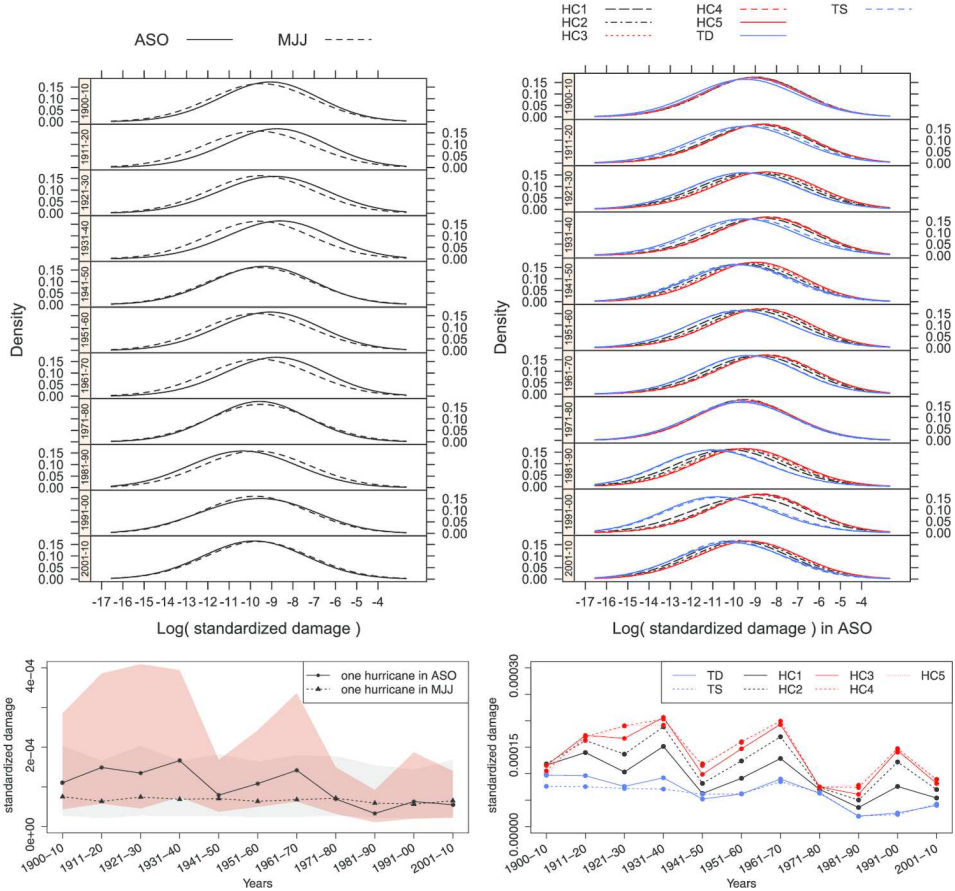}

\caption{Top left panel: the density of logarithmic standardized damage
conditional on MJJ (May--June--July) and ASO (August--September--October).
Top right panel: the density of logarithmic standardized damage in ASO
given the seven maximum wind speed categories defined in Table~\protect\ref{tab:category}.
Bottom left panel: Posterior expectation and 95\% interval (dark gray band
for ASO; light gray band for MJJ) for the median standardized damage of one
hurricane in MJJ and ASO. Bottom right panel: Posterior expectation
for the median standardized damage in ASO for the seven maximum
wind speed categories.}
\label{fig:damage_month}
\end{figure}

Figure~\ref{fig:damage_month} (top left panel) shows the
density of standardized damages (on the log scale) conditional on the
early season and the peak season. The densities of standardized damages
in MJJ (May--June--July) are quite similar throughout all decades, while the
densities in ASO show a moderate decreasing trend across decades.
Figure~\ref{fig:damage_month} (bottom left panel) plots point and
interval estimates for the median standardized damage in the original
scale. From 1900 to 1940, the estimated median standardized damage of
one hurricane in ASO is around twice as large as that in MJJ. However,
from 1941 to 2010, the median standardized damage in ASO depicts
significant variability, with some indication of a slight decreasing
trend across decades.
These results are similar to the ones reported in \citet{Katz2002} and
\citet{Pielke2008}, based on essentially the same data set, albeit
under different damage normalization methods. In particular,
\citet{Katz2002} normalizes the damage during 1925--1995 to 1995 values
and uses a log-normal distribution to fit the damage of individual
storms, finding only weak evidence of a trend in the median of
log-transformed damage. Likewise, in \citet{Pielke2008} hurricane
damage is normalized to 2005 values. In this case, the conclusion is
that there is no long-term increasing trend in hurricane damage
during the 20th century, once societal factors are removed.
We also note here that \citet{Eric2011} detected a significant negative
trend in hurricane damage. Their results are based on the same damage
standardization method as the one we use, but for a different data
set comprising hurricane damages from 1980--2009 in the U.S. and
Canada.

The right-hand side panels of Figure~\ref{fig:damage_month} focus
on the analysis of damage, conditional on the seven different types of
hurricanes that occurred during ASO. The top panel reports the
densities for logarithmic standardized damage conditional on the
different hurricane categories. The bottom right panel
reports the posterior expectations for the corresponding median
standardized damage. Overall, we observe that the higher the category,
the larger the standardized damages tend to be.
Standardized damages were very similar for the
hurricanes recorded in ASO of decade 1971--1980, which is reflected in
both types of inference shown in Figure~\ref{fig:damage_month}.
Standardized damages for TDs and TSs have indistinguishable
distributions. Likewise, at the opposite end of the scale,
damages due to HC4 and HC5 hurricanes are very similar.
This is also due to the data sparseness of TDs and HC5 hurricanes
(only 4 TDs and 3 HC5 hurricanes).

\begin{sidewaysfigure}

\includegraphics{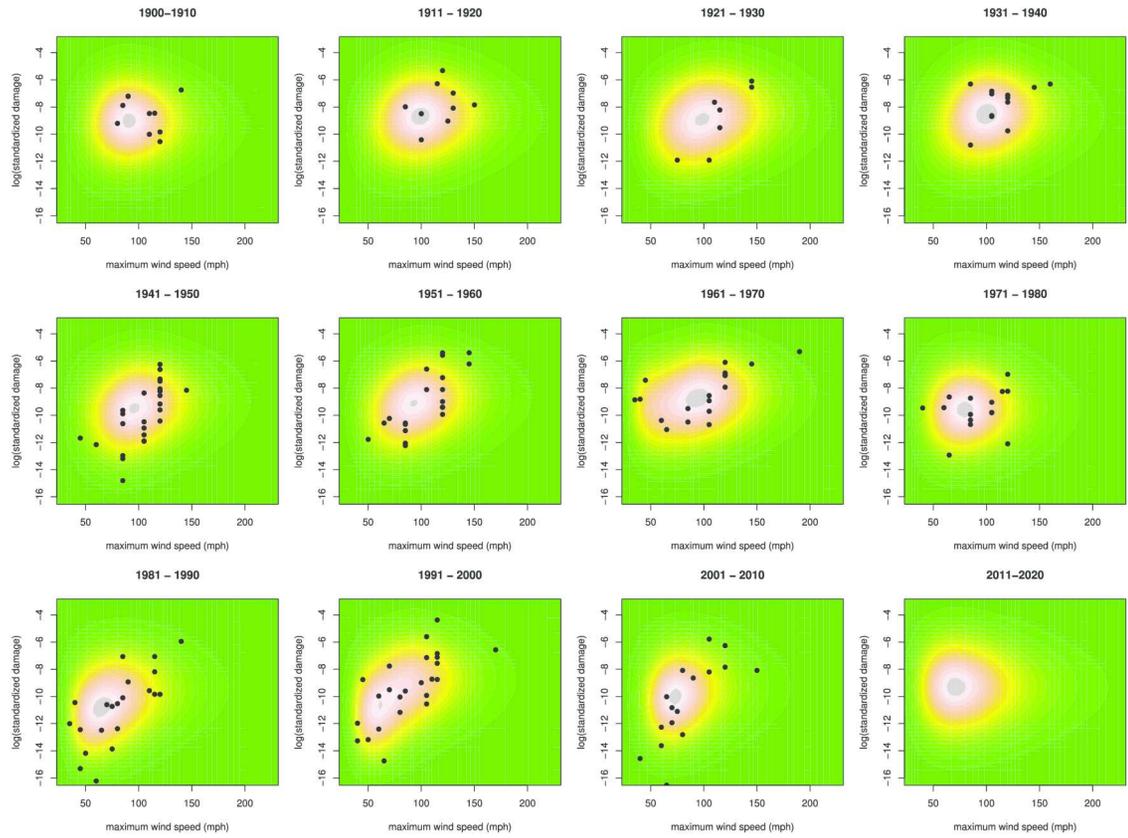}

\caption{Bivariate densities of maximum wind speed (mph) ($x$-axis)
and logarithmic
standardized damage ($y$-axis) in ASO across decades.
The dots correspond to observations in ASO.}
\label{fig:bimark}
\end{sidewaysfigure}

Figure~\ref{fig:bimark} presents the bivariate densities of maximum wind
speed and logarithmic standardized damage given the ASO period, for each
decade. The last panel corresponds to the forecast density for 2011--2020.
We note that only a handful of observations correspond to ASO
in each particular decade. Thus, the results in Figure~\ref{fig:bimark}
are possible owing to our model's ability to borrow strength from all
the available data.
Noteworthy are the positive association between maximum wind
speed and damage after the third decade, and the changes in the
density shapes across the decades, especially 1961--1970 and 1991--2000.
We also note the decrease in maximum wind speeds, starting in 1961--1970.
Overall, from 1961, both the maximum wind speed and standardized
damage have a general decreasing trend. This is a reflection of the
fact that fewer hurricanes with extremely high maximum wind speed
have occurred in recent decades.
Regarding previous related work, \citet{MurnaneElsner2012} modeled the
relationship between wind speed and normalized economic loss as
exponential through quantile regression methods, using all hurricanes
in the 20th century. Our methodology allows for a more comprehensive
investigation of the relationship between hurricane damage and maximum wind
speed, in particular, it enables study of its dynamic evolution across
decades, without the need to rely on specific parametric regression forms.

\begin{figure}

\includegraphics{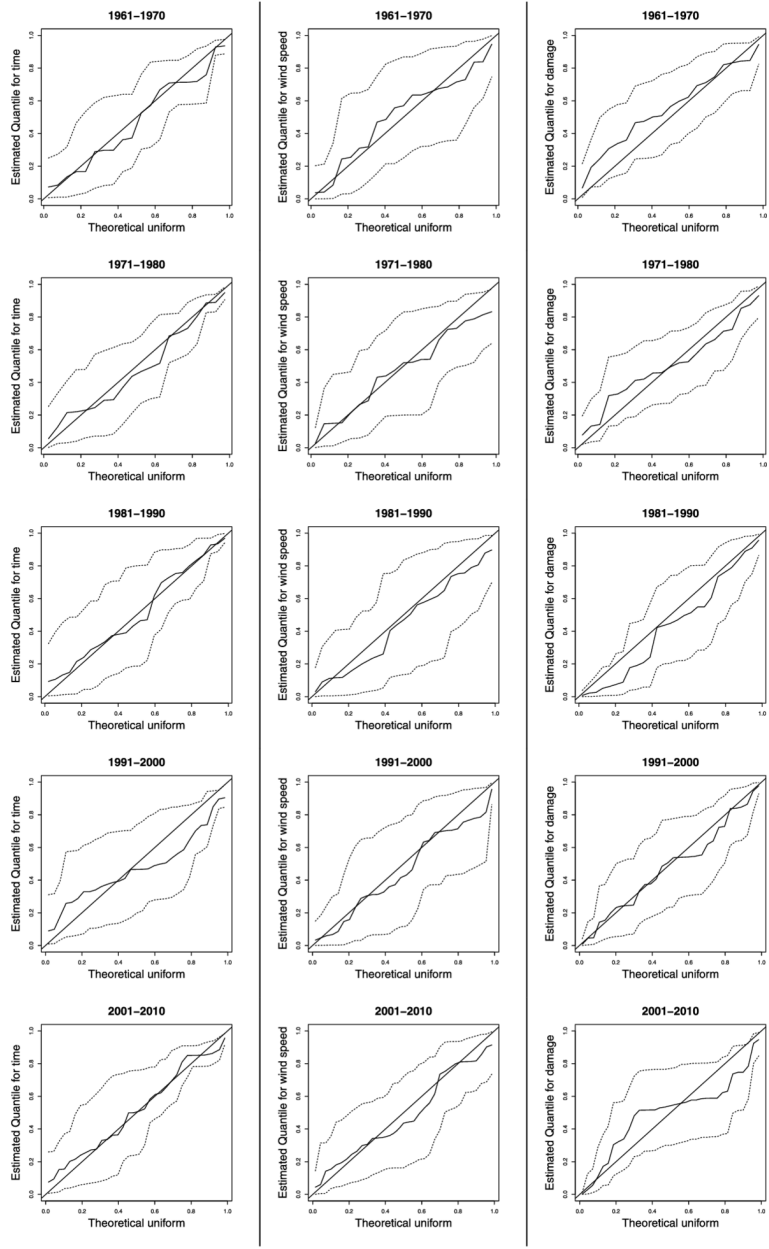}

\caption{Posterior Q--Q plots (mean and 95\% interval) of estimated quantiles
against the theoretical $\operatorname{uniform}(0,1)$ for: time (left panel), maximum
wind speed given time (middle panel), and standardized damage given
time (right panel). Results are shown for the last five decades.}
\label{fig:modelcheck}
\end{figure}

\subsection{Model assessment}

The modeling approach is based on the assumption of a NHPP over the
joint marks-points space. To check the\break NHPP assumption, we use the
Time-Rescaling theorem [Daley and\break Vere-Jones (\citeyear{Daley2003})], according to which,
in each decade, the cumulative intensities
between successive (ordered) observations, $\{ \gamma_k
\int_{t_{i-1,k}}^{t_{i,k}} f_{k}(t) \,dt \}$, are independent exponential\vspace*{-1pt}
random variables with mean one. Thus, $\{ 1 - \break \exp ( -
\gamma_k \int_{t_{i-1,k}}^{t_{i,k}} f_{k}(t) \,dt  ) \}$ are
independent $\operatorname{uniform}(0,1)$ random variables. Likewise, the Poisson
process assumption for the marks implies that the sets of random
variables defined by the c.d.f. values of the conditional mark
distributions, $\{ F_k(y_{i,k} \mid t_{i,k}) \}$ and $\{ F_k(z_{i,k}
\mid t_{i,k}) \}$, are independent $\operatorname{uniform}(0,1)$ random variables.
Hence, the NHPP assumption over both time and marks can be checked by
using the MCMC output to obtain posterior samples for each of the three
sets of random variables above, in each decade. Figure~\ref{fig:modelcheck} shows the Q--Q plots of estimated quantiles for
time, maximum wind speed and standardized damage versus the theoretical
uniform distribution, for the last five decades. The results seem
acceptable, especially in consideration of the limited sample sizes in
each decade.

\begin{figure}

\includegraphics{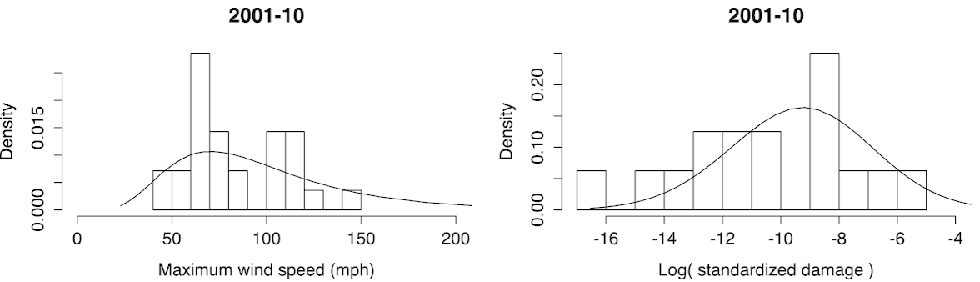}

\caption{Cross-validated posterior predictive densities in ASO of
decade 2001--2010: the left panel corresponds to maximum wind speed,
and the right panel to logarithmic standardized damage. The histograms
plot the associated observations in ASO of 2001--2010.}\vspace*{5pt}
\label{fig:postpred}
\end{figure}

As discussed earlier, Figures~\ref{fig:decade8-11} and \ref{fig:boxplot}
provide visual goodness-of-fit evidence for the model on hurricane
occurrences, by comparing different types of model-based inferences to
the corresponding observations. Similar evidence is provided in
Figure~\ref{fig:bimark} for the maximum wind speed and log-damage relationship.
We also explored other functionals of the model, obtaining similar
results. In addition, we performed posterior predictive checks to
study the model's ability to predict the marks in the 11th decade,
based on the data of the previous 10 decades. In particular, we
implemented the model using only the 204 hurricanes from
1900--2000, and obtained the posterior predictive density of
maximum wind speed and logarithmic standardized damage in ASO of
the 11th decade (2001--2010). Figure~\ref{fig:postpred} shows the
posterior predictive densities superimposed on the histograms of
corresponding observations in ASO of 2001--2010. The histogram in the
left panel corresponds to 28 hurricanes, whereas the one in the right panel
corresponds to only 16 hurricanes, since the damages of the other 12
hurricanes are missing. We notice that the predictions are fairly
compatible with the cross-validation data.

\section{Conclusion}\label{sec5}

We have developed a Bayesian nonparametric modeling method for
seasonal marked point processes and applied it to the analysis of
hurricane landfalls with reported damages along the U.S. Gulf and
Atlantic coasts from 1900 to 2010. Our basic assumption is that
hurricane occurrences follow a nonhomogeneous Poisson process,
with the focus on flexible modeling for dynamically evolving
Poisson process intensities.
The proposed DDP-PBAR model builds from a DDP mixture prior
for the normalized intensity functions based on a PBAR process for
the time-varying atoms, and a parametric time-varying model for
the total intensities. Inference for different Poisson process
functionals can be obtained by MCMC posterior simulation. To
incorporate time-varying marks into the inferential framework for
our motivating application, we have extended the DDP-PBAR mixture model
by adding DDP-AR components for maximum wind speed and economic
damages associated with each hurricane occurrence.

In the analysis of the hurricane data, we have used aggregation to
study the dynamic evolution of hurricane intensity over decades.
The model uncovers different shapes across decades which, however,
share common features with respect to the off-season in May and June
and the peak month of September. The results indicate an increase in
the number of landfalling hurricanes and a decrease in the median
maximum wind speed at the peak of the season across decades.
In the off season, both the number of hurricanes and the maximum
wind speed show little variation across decades.
To study economic loss as a mark, we have introduced standardized
damage to adjust hurricane damages such that they are comparable both
in time and space. We found a slight decreasing trend in
standardized damage of hurricanes in the peak season,
which is also present conditional on the distinct hurricane categories.

With respect to the scientific context of the motivating application, our
work provides a general framework to tackle different practically
relevant problems. The key distinguishing feature of our approach
relative to existing work involves the scope of the stochastic
modeling framework under which the various inferences are obtained.
As discussed in the \hyperref[sec1]{Introduction}, current work is limited to either
estimating trends in hurricane occurrences at the annual level or
estimating the hurricane intensity based on the fully aggregated data,
thus ignoring dynamics across years. Moreover, when incorporating
information on marks, existing approaches oversimplify the underlying
point process structure by imposing homogeneity for the hurricane intensity.
These assumptions are suspect, as demonstrated with the
exploratory data analysis of Section~\ref{section2}.
The proposed Bayesian nonparametric methodology enables flexible
estimation of dynamically evolving, time-varying hurricane intensities
within each season, and therefore has the capacity to capture trends
during particular periods within the hurricane season.
The full inferential power of the modeling framework is realized with
the extension to incorporate marks, which are included as random
variables in the joint model rather than as fixed covariates
as in some of the previous work.
From a practical point of view, the key feature of the model for the point
process over the joint marks-points space is its ability to provide
different types of general conditional inference, including full
inference for dynamically evolving conditional mark densities given
a time point, a particular time period and even a subset of marks.

In summary, the focus of this paper has been in developing a model
that can quantify probabilistically the inter-seasonal and intra-seasonal
variability of occurrence of a random process and its marks, jointly
and without restrictive parametric assumptions. The model is
particularly well suited for the description of irregular long-term
trends, which may be present in the observations or in subsets of the
records. To enhance the forecasting ability of the model, future work
will consider extensions to incorporate external covariates (such as
pre-season climate factors) in a similar fashion to \citet{Katz2002},
\citet{JaggerElsnerBurch2011}, and \citet{ElsnerJagger2013}, albeit under
the more general statistical modeling framework developed here.

%
%
\begin{appendix}
\section*{Appendix: MCMC algorithm for the DDP-PBAR model}\label{app}
\label{appendix}

The DDP-PBAR model for the data $\{ t_{i,k} \}$ can be expressed as follows:
\begin{eqnarray*}
 t_{i,k} \mid G_k, \tau&\sim&\int\operatorname{Beta}\bigl(
\mu\tau, (1-\mu)\tau\bigr) \,dG_k(\mu),\qquad i=1, \ldots, n_k;
k=1,\ldots, K,
\\
 G_k(\mu) &=& \sum_{j=1}^N
w_j \delta_{\mu_{j,k}}(\mu),
\\
 z_j &\sim&\operatorname{Beta}(1,\alpha),\qquad w_1 =
z_1; \\
 w_j &=& z_j \prod
_{r=1}^{j-1} (1- z_r),\qquad j=1,\ldots, N-1;\\
w_N &=& 1-\sum_{j=1}^{N-1}
w_j,
\\
 \mu_{j,k} &= & v_{j,k} u_{j,k} \mu_{j, k-1} +
(1- v_{j,k}), \\
v_{j,k} &\sim&\operatorname{Beta}(1, 1-\rho),\qquad
u_{j, k} \sim\operatorname {Beta}(\rho, 1-\rho).
\end{eqnarray*}
We use an MCMC algorithm to draw posterior samples of
$(\{\mu_{j,k}\}, \{v_{j,k}\}, \break \{ w_j \},  \rho, \alpha)$, including
blocked Gibbs
sampling steps for the DDP prior parameters [\citet{Ishwaran2001}].
Configuration variables $\{L_{i,k} \} $ are introduced to indicate
the mixture component to which each observation is allocated.
We use $n^*$ to denote the number of distinct values in the $\{ L_{i,k}
\}$,
and $L^*= \{ L_{j}^*\dvtx j=1,\ldots,n^* \}$ for the set of distinct values.

The first step is to update the atoms $\{ \mu_{j,k} \}$, which depends
on whether $j$ corresponds to an active component or not.
When $j \notin L^*$, $\mu_{j,1} \sim\operatorname{Unif}(0, 1)$, and for $k=2,\ldots,K$,
$\mu_{j,k}$ is drawn from $p(\mu_{j,k} \mid\mu_{j,k-1},
v_{j,k},\rho)$,
which is a scaled Beta distribution arising from the PBAR process,
\[
p(\mu_{j,k} \mid\mu_{j,k-1}, v_{j,k},\rho) =
\frac{1}{v_{j,k} \mu_{j,k-1}} \operatorname{Beta}\biggl(\frac{\mu_{j,k} + v_{j,k} - 1}{v_{j,k} \mu_{j,k-1}} \Big\mid\rho, 1-\rho
\biggr),
\]
where
$\mu_{j,k} \in ( 1- v_{j,k} , \min\{ 1, 1 -v_{j,k} + v_{j,k}
\mu_{j,k-1} \}  )$.
When $j \in L^*$, the posterior full conditional for $\mu_{j,1}$ is
proportional to $\prod_{i=1, \{L_{i,1}=j\}}^{N_1}
\operatorname{Beta}(t_{i,1} \mid\mu_{j,1} \tau, (1-\mu_{j,1}) \tau) p(\mu
_{j,2} \mid\mu_{j,1}, v_{j,2}, \rho) p(\mu_{j,1})$.
For $k=2,\ldots, K-1$, the full conditional for $\mu_{j,k}$ is
proportional to
$\prod_{i=1, \{L_{i,k}=j\}}^{N_k} \operatorname{Beta}(t_{i,k} \mid\mu_{j,k}
\tau, (1-\mu_{j,k}) \tau)
p(\mu_{j, k+1} \mid\mu_{j,k},\break   v_{j,k+1}, \rho) p(\mu_{j,k} \mid
\mu_{j, k-1}, v_{j,k}, \rho)$.
Finally, the full conditional for $\mu_{j,K}$ is proportional to
$\prod_{i=1, \{L_{i,K}=j\}}^{N_K} \operatorname{Beta}(t_{i,K} \mid\mu_{j,K}
\tau, (1-\mu_{j,K}) \tau)
p(\mu_{j,K} \mid\mu_{j,K-1}, v_{j,K}, \rho)$.
We use Metropolis--Hastings steps to update the $\mu_{j,k}$, with the
proposal distribution taken to be $p(\mu_{j,k} \mid\mu_{j, k-1},
v_{j,k}, \rho)$.

The sampling of weights $\{ w_{j} \}$, configuration variables $\{
L_{i,k} \}$ and
$\alpha$ can be implemented using standard updates under the blocked Gibbs
sampler. Updating the latent variables $\{ v_{j,k} \}$ involves only
the PBAR process.
The full conditionals are given by
\begin{eqnarray*}
&& p(v_{j,k} \mid\mu_{j,k}, \mu_{j, k-1}, \rho)\\
&&\qquad \propto
\frac{1}{v_{j,k} } \operatorname{Beta}\biggl(\frac{\mu_{j,k} + v_{j,k} -
1}{v_{j,k} \mu_{j,k-1}} \Big\mid \rho, 1-\rho
\biggr) \operatorname{Beta}(v_{j,k} \mid1, 1-\rho),
\end{eqnarray*}
where $v_{j,k} \in ( 1-\mu_{j,k} , \min\{ 1,\frac{1-\mu
_{j,k}}{1-\mu_{j,k-1}} \}  )$,
and sampling from each of them was implemented with a Metropolis--Hastings
step based on $\operatorname{Beta}(1, 1-\rho)$ as the proposal distribution.
Finally, the PBAR correlation parameter $\rho$ is also sampled using a
Metropolis--Hastings step.
\end{appendix}

\section*{Acknowledgments} The authors wish to thank Roger Pielke and
Kevin Sharp for helpful
discussions regarding the hurricane data, as well as the Editor,
Tilmann Gneiting, an Associate Editor and a referee for their
constructive feedback.

%
%

%

%



\printaddresses
\end{document}